\documentclass[conference]{IEEEtran}

\IEEEoverridecommandlockouts
\usepackage{cite}
\usepackage{amsmath,amssymb,amsfonts}
\usepackage{algorithmic}
\usepackage{textcomp}
\usepackage{xcolor}
\usepackage{graphicx}
\usepackage{booktabs}
\usepackage{xcolor}
\usepackage{url}

\usepackage[breaklinks]{hyperref}
\usepackage{cleveref}
\usepackage{multirow}
\usepackage{array}
\usepackage{tabularx}
\usepackage{float}

\newcolumntype{L}{>{\raggedright\arraybackslash}X}

\def\BibTeX{{\rm B\kern-.05em{\sc i\kern-.025em b}\kern-.08em
    T\kern-.1667em\lower.7ex\hbox{E}\kern-.125emX}}

\begin{document}

\title{An Engineering Journey Training Large Language Models at Scale on Alps: The Apertus Experience}


\author{
\IEEEauthorblockN{
Jonathan Coles\IEEEauthorrefmark{1}, Stefano Schuppli\IEEEauthorrefmark{1}, Lukas Drescher\IEEEauthorrefmark{1}, Fawzi Roberto Mohamed\IEEEauthorrefmark{1}, Elia Palme\IEEEauthorrefmark{1}, Henrique Mendonça\IEEEauthorrefmark{1},\\
Miguel Gila\IEEEauthorrefmark{1}, Mark Klein\IEEEauthorrefmark{1}, Maxime Martinasso\IEEEauthorrefmark{1}, Joost VandeVondele\IEEEauthorrefmark{1}, Torsten Hoefler\IEEEauthorrefmark{1}, Thomas Schulthess\IEEEauthorrefmark{1},\\
Josh Romero\IEEEauthorrefmark{2}, Igor Gorodetsky\IEEEauthorrefmark{3}, Ryan Hankins\IEEEauthorrefmark{3}, Isa Wazirzada\IEEEauthorrefmark{3},\\
Martin Jaggi\IEEEauthorrefmark{4}, Antoine Bosselut\IEEEauthorrefmark{4}, Imanol Schlag\IEEEauthorrefmark{5}, Antoni-Joan Solergibert i Llaquet\IEEEauthorrefmark{4}, Alejandro Hernández Cano\IEEEauthorrefmark{4},\\
Theofilos Ioannis Manitaras, Nicholas John Browning
}

\IEEEauthorblockA{\IEEEauthorrefmark{1}Swiss National Supercomputing Centre (CSCS)}
\IEEEauthorblockA{\IEEEauthorrefmark{2}NVIDIA}
\IEEEauthorblockA{\IEEEauthorrefmark{3}HPE}
\IEEEauthorblockA{\IEEEauthorrefmark{4}EPFL}
\IEEEauthorblockA{\IEEEauthorrefmark{5}ETH Zurich}
}

\maketitle

\begin{abstract}
Large Language Models (LLMs) have surged as a transformative technology for science and society, prompting governments worldwide to pursue sovereign AI capabilities that ensure data compliance and cultural representation. However, the associated capital costs and engineering complexity required to train these models have largely restricted such capabilities to the private sector, leaving a significant gap for public institutions. 

This paper details the engineering journey behind training \textit{Apertus}, a fully open multilingual foundation model, on the \textit{Alps} supercomputer. Representing a first-of-its-kind achievement for academia at the 70B parameter scale, we successfully deployed a massive pre-training campaign on one of Europe's largest systems for open science, powered by NVIDIA GH200 Grace Hopper Superchips. We detail the challenges encountered in readying HPC infrastructure for training AI models, from overcoming storage bottlenecks to stabilizing large-scale interconnects, and the lessons learned in transforming a supercomputer into a resilient software-defined Machine Learning Platform. 

Finally, we discuss the post-training requirements and evolution of our Machine Learning platform, outlining how this initial release lays the groundwork for a sustained, iterative operational capability, in particular for fine tuning foundation models, that extends well beyond a single model training run.
\end{abstract}

\section{Introduction}
\label{sec:intro}

The advent of transformer-based architectures \cite{vaswani2017} has opened a new era of Artificial Intelligence (AI) research, shifting the paradigm from task-specific models to general-purpose Foundation Models. Large Language Models (LLMs) have demonstrated transformative potential not only for societal applications but also as instruments for scientific discovery \cite{ai4science2023,wang2023scientific,AI4Science2023TheIO}. The capabilities of these models have been shown to follow distinct neural scaling laws \cite{kaplan2020, hoffmann2022}, which requires datasets measured in trillions of tokens and compute budgets that span exaflops of low-precision floating-point operations. Consequently, the development of state-of-the-art LLMs has triggered a global infrastructure race. Both nations and hyperscalers have announced data centers with 100,000 GPU scale and power needs approaching the gigawatt range \cite{semianalysis2024}. 

While the private sector has led early development \cite{gpt4_2023, llama3_2024}, there is a critical need for open, transparent, and sovereign models to ensure scientific reproducibility and data compliance. In Europe, this necessity has crystallized into initiatives such as EuroLLM and the Swiss AI Initiative to develop fully open models respecting the EU AI act regulation.
Switzerland has positioned itself at the forefront of this open science movement through the collaboration of EPFL, ETH Zurich, and the Swiss National Supercomputing Centre (CSCS), as well as various other institutions. This effort relies on the Alps Research Infrastructure, a supercomputer featuring NVIDIA GH200 Grace Hopper Superchips \cite{Alps_aVersatileRI}. However, training foundation models on a supercomputer requires a transition from high-performance computing (HPC) workloads to massive-scale data-intensive long-running training campaigns.

The lighthouse of this initiative is \textit{Apertus}\cite{apertus_report_2025}, a fully open LLM released in September 2025. Unlike ``open weights" releases sometimes provided by the private sector, Apertus provides full transparency regarding data pipelines, training code, and evaluation metrics. The training campaign, initiated in early 2025, represented a massive engineering challenge for an academic supercomputing center. The production run for the largest model spanned roughly 3 months, consuming over 6 million GPU hours and sustaining utilization across thousands of GPUs.

However, the successful completion of this campaign was far from guaranteed. Unlike commercial hyperscalers who operate homogeneous, single-purpose AI clusters, academic supercomputing centers like CSCS must support diverse scientific workloads on general-purpose infrastructure. Deploying a massive, fault-intolerant AI model training workload on the Alps system required additional engineering efforts to build a resilient Machine Learning (ML) Platform. This paper details this engineering journey of adapting a leadership-class supercomputer for large-scale AI foundation model training. The specific contributions of this work are:

\begin{enumerate}
    \item \textbf{Pre-Training Requirements:} We outline the preparatory steps required to transform an HPC system into an ML Platform, including storage dimensioning and environment standardization.
    \item \textbf{Operational Challenges:} We classify the obstacles encountered during the multi-month campaign, specifically addressing the ``Data Bottleneck" and the ``Communication Wall" inherent to distributed training at scale.
    \item \textbf{Performance and Stability:} We present empirical results from the campaign, discussing failure modes, recovery strategies, and the reality of having large-scale long running jobs.
    \item \textbf{Infrastructure Lessons:} We provide a discussion from the infrastructure provider's perspective, offering a blueprint for other centers aiming to support large-scale AI workloads.
\end{enumerate}

\section{Infrastructure and Platforms}
\label{sec:infra_and_platforms}
The Apertus training campaign was executed on the \textit{Alps} Research Infrastructure (RI), hosted at CSCS. As the flagship system in the Swiss national HPC strategy, Alps represents a departure from monolithic supercomputing architectures, offering a software-defined infrastructure that allows research groups to tailor the service capabilities to their specific scientific domains.

\subsection{The Alps Research Infrastructure}
\label{sec:alps_ri}

Physically, Alps is built on the HPE Cray EX architecture, leveraging the Slingshot-11 interconnect to unify a heterogeneous set of compute resources via a single high-speed fabric. The system accommodates diverse node architectures to support varying workload requirements, including AMD EPYC ``Rome" CPUs, NVIDIA A100 GPUs, and AMD MI250 and MI300A accelerators.

The center piece of the infrastructure, and the primary resource for Apertus, is the large-scale NVIDIA Grace Hopper partition. This partition aggregates 10,752 GH200 Superchips packaged into 2,688 quad-GPU nodes, capable of delivering approximately 434 PFlops of FP64 performance. Supporting this compute capability is a tiered storage architecture designed to handle diverse I/O patterns:
\begin{itemize}
    \item Performance Scratch: A 5 PB ClusterStor E1000 all-flash filesystem for high-IOPS workloads.
    \item Capacity Scratch: A 100 PB ClusterStor filesystem based on HDD technology.
    \item Service: A 1 PB VAST Data platform for user home directories, augmented by a recent deployment of an additional 10 PB VAST system to support scratch data management.
\end{itemize}

\subsection{The Software-Defined Operating Model}
\label{sec:software_stack}

The transformation of Alps from a hardware installation into a flexible Research Infrastructure is achieved through its novel management software. Recognizing that the traditional ``one-size-fits-all" vendor software stack is increasingly insufficient for modern, diverse scientific workflows, CSCS developed the \textit{Versatile Software-Defined Cluster} (vCluster) technology~\cite{vcluster}.

\begin{figure}[ht]
    \centering
    \includegraphics[width=0.8\linewidth]{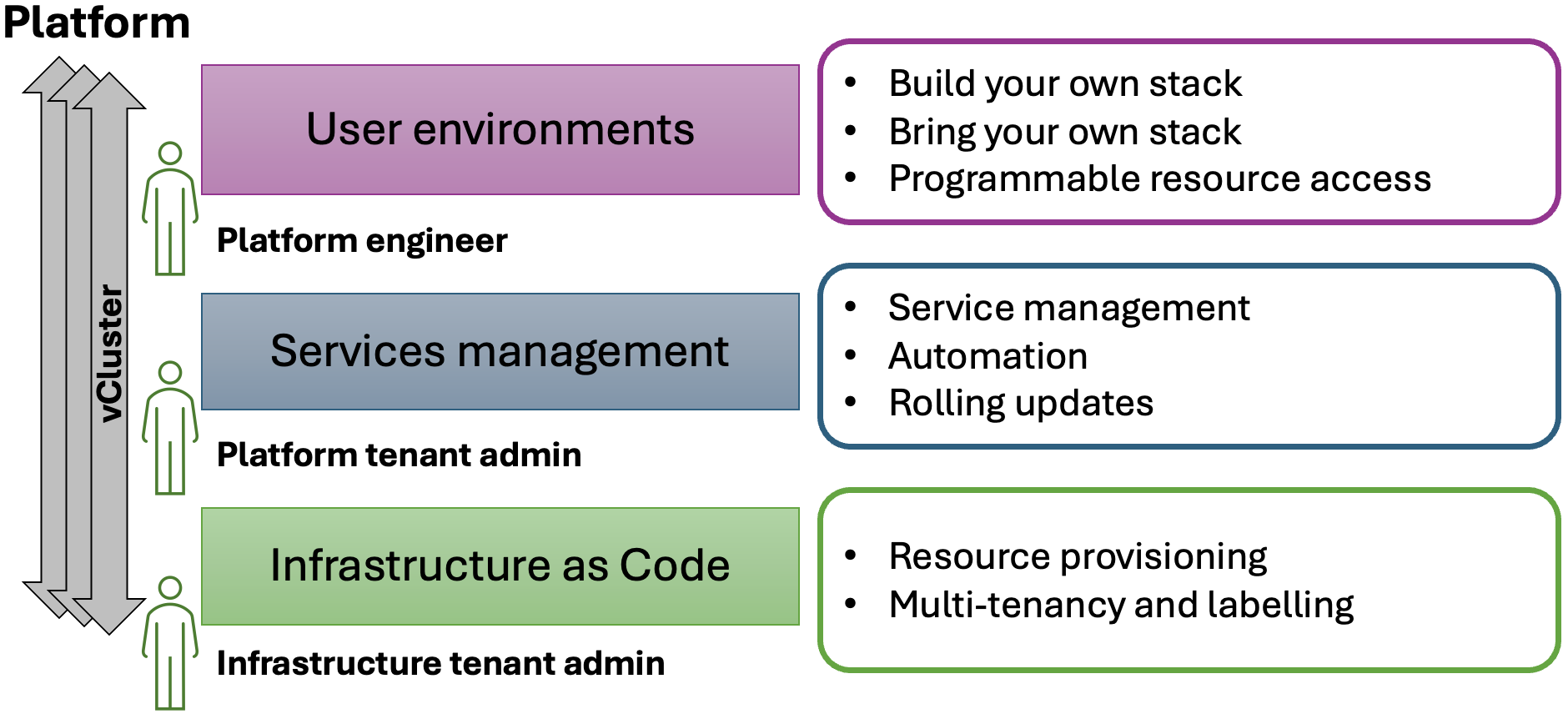}
    \vspace{-0.2cm}
    \caption{The three layers concept of the vCluster technology}
    \label{fig:vcluster}
\end{figure}

This technology adopts a cloud-native approach, decoupling the system into three distinct operational layers as displayed in \autoref{fig:vcluster}:
\begin{enumerate}
    \item Infrastructure as Code Layer (IaC): Manages the bare-metal provisioning and grouping of resources (e.g., booting nodes with minimalistic OS images).
    \item Service Management Layer: Orchestrates the configuration of these node groups, deploying the necessary daemons, packages and configuration required for specific scientific domains.
    \item User Environments Layer: Defines the user-facing interface by composing various ``vServices", such as the Slurm workload manager, container engines~\cite{sarus}, or the FirecREST web-facing API~\cite{firecrest}.
\end{enumerate}
This architecture allows Alps to host multiple, logically isolated platforms simultaneously. Operational since early 2024, the system currently supports distinct platforms for traditional HPC simulations, Climate \& Weather prediction, and, in particular for this work, the ML platform. 

A key advantage of this architecture is resource elasticity. The boundary between platforms is logical rather than physical; nodes can be dynamically reassigned from one platform to another platform by modifying grouping labels in the Infrastructure as Code layer and triggering a reconfiguration of those nodes at the Service Management layer. While this elasticity is currently still a manual operational process, it was instrumental during the Apertus campaign, allowing us to temporarily expand the amount of resources to accelerate training throughput to meet release deadlines. It furthermore allowed us to rapidly iterate through system-level changes to diagnose and address performance issues without disturbing the other platforms.

\subsection{The Machine Learning Platform}
\label{sec:ml_platform}

The development of the ML Platform on Alps has so far been driven by the specific requirements of the Swiss AI Initiative. This program represents the largest open-science effort for AI foundation models worldwide and serves as the inaugural flagship project of the Swiss National AI Institute, a strategic partnership between the ETH AI Center and the EPFL AI Center. 

This community’s computational needs differ fundamentally from those of traditional HPC domains. Beyond raw floating-point performance, Swiss AI researchers require an infrastructure that supports rapid prototyping, massive-scale data ingestion, and the flexibility to modify the software stack down to the driver level. It was to meet the specific requirements of this initiative, and specifically the Apertus training campaign, that the ML Platform was created.

To address the unique requirements of the Swiss AI community, the ML Platform was architected as a composite service consisting of two primary vClusters, each optimized for a distinct phase of the model lifecycle:

\begin{itemize}
    \item \texttt{clariden}: This partition serves as the primary computational engine for the Swiss AI Initiative. Comprising a significant amount of the available NVIDIA GH200 Grace Hopper nodes, Clariden is orchestrated via the Slurm workload manager. It provides a dedicated environment for the development, pre-training, and fine-tuning of foundation large models like Apertus.
    \item \texttt{bristen}: This is a flexible auxiliary partition which, during the Apertus training campaign, ran as a Slurm cluster with legacy NVIDIA A100 GPUs and x86 CPUs. This partition was primarily used for data pre-processing, evaluation, and inference workloads, which typically involve less parallelism but still demand high throughput. It is now being migrated to Kubernetes to support diverse workloads, including long-running inference and auxiliary services that benefit from cloud-native orchestration.
\end{itemize}

\subsection{AI Software Stack Support}
\label{subsec:ai_sw_stack_based_on_containers}

The ML ecosystem evolves rapidly and is highly diverse, with users relying on a constantly changing mix of libraries, frameworks, and tools. To support rapid adoption of new technologies, we let users bring their own software stack instead of promoting centrally maintained environments.

This approach leverages the community’ familiarity with application containers, readily-available vendor-optimised images for established libraries such as PyTorch, and long-standing container developments at CSCS~\cite{sarus} that integrate seamlessly with our HPC ecosystem. Together, these provide a mature, low-friction foundation for ML workloads on Alps.

Achieving bare-metal performance with portable container images requires bridging the isolated container namespace and the HPC hardware. To enable direct access to host-level resources, specifically the GPUs and the Slingshot interconnect, we leverage the Open Container Initiative (OCI) runtime specification. Through OCI lifecycle hooks, the platform intercepts container instantiation to inject host libraries (such as libfabric and CUDA drivers), mount device files, and configure environment variables for the interconnect. A declarative, TOML-based \textit{Environment Definition File} (EDF)~\cite{sarus_EDF} abstracts this from users, enabling high-performance features without managing driver mappings or low-level configurations manually. The Apertus training relied extensively on containers.

To maintain stable throughput, we built performance-monitoring test suites to detect degradations from system or configuration changes and incompatibilities from new software versions. Distributed ML workloads like Apertus span many software layers (\autoref{tab:apertus_layered_stack}), from top-level Python code down to Slingshot drivers, creating complex version interactions that are hard to test exhaustively. In practice, the community relies on a set of well-known base images (e.g., NGC’s PyTorch \cite{nvidia_ngc_catalog}), so we focused system-level monitoring on a representative subset while still capturing the most relevant performance and compatibility issues.

\section{Challenges Ahead of Training an LLM}
\label{sec:challenges}

Training a foundational LLM follows an evolving workflow defined by research teams and enabled by infrastructure and services. It spans typical stages such as dataset curation and preparation, model and training recipe design, large-scale pre-training, successive post-training phases, and downstream evaluation and release.

For infrastructure providers, the primary objective is to deliver compute, storage, and software foundations that enable such workflows to execute reliably at scale, integrate across system layers, and operate with minimal manual intervention. The Apertus campaign was the first instance in which a complete foundational LLM training workflow was exercised at this scale on Alps. This milestone represented a platform maturation phase: while individual components of the system were already production-grade, their simultaneous use under sustained, LLM-scale load exposed operational and integration challenges that are non-trivial to anticipate in isolation.

Addressing these challenges required focused engineering effort, close coordination with users, and iterative refinement of operations, monitoring, and service interfaces. The resulting experience, some improvements now integrated into the standard service, others informing ongoing and planned work, is distilled here into a set of representative observations drawn from preparatory activities, intended to guide future infrastructure and service development for large-scale AI workloads.

\subsection{Data Acquisition Through Web Crawling}

Most Apertus pre-training datasets were obtained via conventional methods, such as downloads from established archives or direct data transfers.

During the project, however, we observed large-scale web crawling activities (e.g., to verify opt-out preferences fetching up-to-date robots.txt files) launched from within the CSCS environment. The resulting traffic triggered alerts from external Internet Service Providers (ISPs): requests from CSCS public IP ranges resembled distributed-denial-of-service (DDoS) attacks, since the available external network bandwidth of a datacentre like CSCS (400~Gbit/s) easily creates a significant load for an ISP. Blacklisting of our public IPs (e.g., at the DNS level) would have disrupted access to external services critical to other scientific workloads also hosted at CSCS. This incident highlighted the need for explicit policies and proactive monitoring to prevent infrastructure-level risks from unintended use of shared HPC systems. We have since updated our procedures and monitoring accordingly.

\subsection{Dataset tokenization}

The full dataset was tokenized before pre-training using a preprocessing pipeline that read Snappy-compressed Parquet shards from Lustre and produced Megatron-compatible \texttt{.bin} and \texttt{.idx} files. To tune the tokenization setup, users varied output shard size, file count, and workers per node, achieving throughputs between 51 and 72 million tokens per second per node. The optimal setup found required about 60 node hours for the full dataset tokenization.

Both the tuning phase and production run placed heavy load on shared storage given the different underlying usage patterns involved. This temporarily reduced filesystem availability for other users and required close coordination with those running the tokenization. The experience showed that tokenization, though seemingly smaller and shorter, can be highly storage-intensive and must be planned and managed as carefully as large-scale training runs.

\subsection{Dataset volume, layout, and storage considerations}

The Apertus team curated the datasets used in the training and stored them in the binary, index-based format commonly used in Megatron-LM and similar large-scale LLM pipelines. Each dataset comprises a large \texttt{.bin} file of tokenized text serialized as contiguous integer sequences, plus a compact \texttt{.idx} file that encodes document boundaries and offsets. This design supports efficient sequential reads and memory-mapped access to large token buffers while enabling flexible document-level indexing with minimal metadata.

CSCS supported the effort by providing advisory and operational support around the scratch Lustre filesystem, required to host and access these datasets at scale. The tokenized datasets occupied about 63 TB on disk across roughly 2'800 large shard files, averaging $\sim$22 GB each. This large-shard design effectively minimised metadata overhead and avoided small-file pressure, and aligned well with the sequential, high-throughput access patterns typical of LLM training workflows.

During training and data processing, many ranks read the shards concurrently, creating a read-intensive workload. Although the filesystem layout supported efficient sequential reads, these patterns caused variable aggregate throughput over long runs. The issue was operational only and did not affect data integrity or training correctness, but required mitigation. As part of broader throughput-stabilisation measures, the datasets were moved to the SSD-based storage system (\autoref{sec:alps_ri}), which offered higher, more predictable read bandwidth under concurrent access and thus more stable training throughput.

An additional effect appeared when moving from the 70B to smaller models like the 8B variant. Because of their higher training throughput, these models generated a higher sustained read load on the filesystem, even with fewer nodes. As a result, infrastructure issues that seemed resolved in the 70B runs reappeared, indicating that Lustre-hosted datasets may need more striping to distribute read pressure. This shows that optimal filesystem tuning depends on model throughput, not just dataset size or model parameters count.

A similar scale-dependent issue arose with application container images. Image loading time, generally negligible, became significant at large node counts. To address this, the squashed container images were explicitly striped, improving aggregate read bandwidth at job launch.

\subsection{Custom activation function implementation}
The model architecture, including the adoption of the xIELU activation function \cite{xIELU_2025} in the MLP layers of the transformer blocks, was defined within the Apertus project team. When executed at scale, a blocking technical issue emerged: for specific container-image versions, the Python reference implementation of xIELU was incompatible with \texttt{torch.compile} \cite{PT2.0_compiled}, causing JIT compilation to fail, preventing execution.

CSCS engineers supported the users by diagnosing the issue within the lower layers of the software stack and developing a custom CUDA kernel that restored compatibility with the compilation pipeline. Building on this intervention, we further analysed the activation’s computational structure and implemented optimisations to improve runtime efficiency, resulting in approximately 20\% kernel execution speedup. This episode illustrates how specialised systems and performance-engineering expertise complements user-driven architectural choices, enabling research-oriented components to be transformed into production-oriented and high-performance implementations.

\subsection{The Training Recipe}

The Apertus training recipe, defined by the scientific teams, spans the model architecture, parallelisation strategy, environment definitions (EDF), launch scripts, and includes the supporting post-training components. The model architecture is implemented on top of NVIDIA’s Megatron-LM \cite{megatronlm}, which provides native support for different parallelism methods. The codebase was forked \cite{apertus_report_2025, saii_github} to incorporate the specific adaptations required, including the xIELU activation function.

Training was orchestrated through user-developed launch scripts that also encoded the parallelisation configuration, batch-size and learning-rate schedules, and check-pointing logic. These scripts executed through the CSCS container engine on user-provided container images, rather than centrally curated environments, and implemented wall-time–aware termination by monitoring job runtime and writing a final checkpoint shortly before allocation expiration. Combined with user-specific Slurm priorities and suitably over-sized partitions, this approach enabled continuation through successive allocations without loss of progress.

Training runs were executed inside curated container images derived from official NVIDIA GPU Cloud (NGC) \cite{nvidia_ngc_catalog} PyTorch bases, encapsulating the dependency stack required by Megatron-LM and the Apertus extensions \cite{saii_github}.

To satisfy model sizes and token budget goals, the Apertus team used a hybrid approach combining data (DP), tensor (TP), pipeline (PP), and, for long sequences, context parallelism (CP). TP was fixed at 4 to match each node’s GPU topology, while DP and PP were tuned per training phase to satisfy memory and scaling needs for the 8B and 70B models, and also to adapt to the available cluster resources. This aligned well with the vCluster-based nodes management on Alps \cite{Alps_aVersatileRI}, where each vCluster can be scaled up or down to meet the evolving needs of each community.

\section{The Journey}
\label{sec:journey}

\begin{figure*}[ht]
    \centering
    \includegraphics[width=0.7\textwidth]{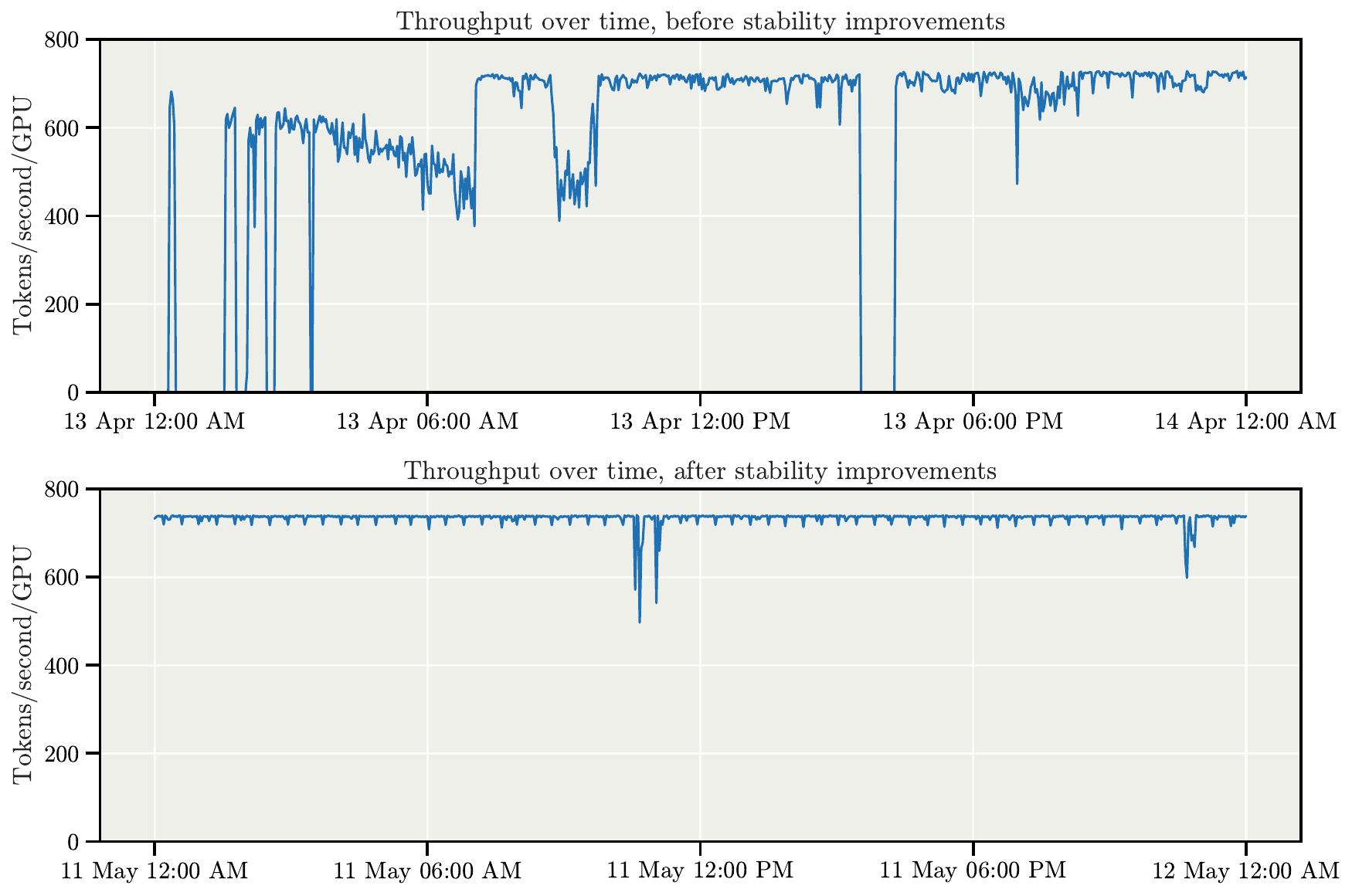}
    \caption{
    \textbf{Comparison of throughput of the 70B Apertus pre-training on 2048 GPUs before and after stability improvements.}
    \textbf{Top:} Runs prior to stability tuning show high variability and frequent restarts, largely driven by poor filesystem I/O before migrating to full-flash storage, and an NVIDIA driver issue related to access counter-based memory page migration.
    \textbf{Bottom:} Performance after stability enhancements, exhibiting consistent throughput with predictable dips corresponding to Python garbage collection and asynchronous checkpointing. Residual irregular fluctuations are attributable to minor filesystem interference.
    Figure and caption text reproduced from the Apertus Technical Report\cite{apertus_report_2025} for clarity.
    }
    \label{fig:TSG-improvements}
\end{figure*}

With the model, data, and software stack in place, the Apertus campaign moved into execution. Scaling training pipelines to sustained production revealed new system-level challenges, discussed in this section. These are organised as follows: section A covers issues across software layers, section B the data bottleneck, section C the communication wall, section D discusses long-running stability topics, and section E enumerates several of the approaches adopted or being developed since the campaign to improve the experience of user and engineering teams alike. The combined positive impact of this engineering work is shown in \autoref{fig:TSG-improvements}.

\subsection{From Sandbox to Production}
\label{subsec:journey__sandbox_to_prod}

Moving the project from development to production, and thus scaling out workloads, exposed new behaviors that hindered jobs from starting or running stably.

\subsubsection{Interplay Between Linux, GPU Drivers, Interconnects, and I/O Subsystems}

Running the Apertus pre-training workload at scales of up to 4096 GPUs exposed tightly coupled issues spanning GPU drivers, the Linux kernel, the Slingshot-11 interconnect, and shared storage (\autoref{tab:apertus_layered_stack}). Early production runs showed substantial performance variability between runs, seen as fluctuations in effective tokens-per-second throughput (\autoref{fig:TSG-improvements}). Analysis revealed that the dominant causes were infrastructure interactions, especially collective communication stability and I/O behavior, rather than algorithmic effects.

After performing a parameter study of ML-related communication collectives, we identified a degradation in performance as we scaled to large node counts. Working closely with our industry partners we ultimately identified the root cause as a subtle issue with libfabric when using the Slingshot alternative \texttt{read} protocol. We found that the restricted \texttt{get} was being transmitted via a default traffic class, which resulted in delays of control traffic due to network congestion \cite{libfabric_pr11079}. Modifying libfabric to cause the restricted \texttt{get} requests to use the appropriate traffic class reduced their latency under congestion. The fix is available in libfabric from version 2.2.

At the interconnect layer, inconsistent combinations of the OFI NCCL plugin \cite{aws_ofi_nccl_2025} and \texttt{libfabric} \cite{libfabric_website} versions led to unstable collective operations and slow or unreliable checkpoint restarts. 
Aligning these components to a validated version set restored predictable communication across many nodes, illustrating how small version mismatches in the communication stack can cause large-scale throughput variability.

Several critical defects were also identified in the interaction between the GPU driver and the Linux kernel. In particular, a race condition in the kernel \cite{lkml_mmu_bug}, reachable through GPU driver calls frequently caused node crashes during long-running jobs and required a targeted fix to stabilise the system. Even though this bug was first reported and patched in 2022, the latest available kernel validated on the system during training was still vulnerable. To allow for fast deployment to nodes, a lightweight workaround to avoid the race-condition was developed by patching the NVIDIA driver module with its own implementation of the kernel function fix. 

The campaign highlighted that default Linux settings are not always suitable for extreme-scale ML workloads: Transparent Huge Pages (THP) \cite{linux_kernel_transhuge}, while beneficial for many general-purpose applications, consistently degraded performance for Apertus and were therefore exposed as a Slurm-level option, allowing users to disable them selectively for sensitive runs.

\subsubsection{Unified Memory and Heterogeneous Memory Effects}

The GH200 architecture \cite{evans2022_gh200, fusco2024_gh200} combines CPU-side LPDDR5X and GPU-side HBM3 into a unified, cache-coherent memory system. At scale, this exposed several non-obvious interactions between the Linux kernel, GPU drivers, and system services.

The Linux kernel caches file-backed data in memory, and on GH200 these OS file caches may be stored in GPU-attached HBM. In principle, such caches should be evicted or migrated back to CPU memory when GPU memory is required. In practice, a kernel–driver defect prevented reliable migration and eviction, leading to persistent file caches occupying GPU memory and increasing the frequency of GPU out-of-memory (OOM) events during training. Due to the non-deterministic placement of cached pages, these failures manifested inconsistently across ranks and nodes. 

As an operational mitigation, file caches were explicitly flushed prior to job start, and a Slurm prolog enforced a pre-flight check requiring at least 90\% of GPU memory to be allocatable before a node could enter a user allocation. While effective at job launch, this approach could not fully prevent cache re-accumulation during long-running workloads. The issue has since been addressed in R570 of the driver.

A further memory issue arose from the GPU driver’s access-counter–based page migration mechanism, introduced in R550, which automatically promotes frequently accessed CPU pages into GPU memory. Until driver version R565, this mechanism generated interrupt storms on specific CPU cores, resulting in highly unpredictable performance and repeated kernel warnings. The feature was therefore disabled at kernel boot until a fixed driver version could be installed.

Additional mitigations constrained system processes, memory allocations, and \texttt{tmpfs} allocations to CPU NUMA nodes, reducing interference between OS activity and user workloads. Together, these measures restored predictable GPU memory availability and stable training behaviour, illustrating that unified memory systems require explicit OS- and driver-level controls to operate reliably.

\subsubsection{Mitigating Runtime Compiler Failures}

During pre-training, intermittent and non-deterministic crashes occurred at job startup. Though not easily reproducible under testing conditions, failure probability increased with job scale. Initially believed to be application issues, our analysis showed the failures originated below the framework layer.

Specifically, the crashes were traced to a defect in \texttt{libnvrtc}, NVIDIA’s runtime compilation library responsible for just-in-time (JIT) compilation of GPU kernels, included in NGC 25.01. The runtime compiler intermittently produced segfaulted. At scale, this caused near-certain job startup failure, despite identical user code and launch configurations.

As an interim mitigation, a custom container image was produced that bundled an updated version of \texttt{libnvrtc}, decoupling the runtime compiler from the default library set provided by the base container environment. This workaround eliminated the startup crashes and enabled stable production runs while upstream fixes were pending.

The issue was subsequently resolved in newer software stacks, and Apertus later upgraded to a newer PyTorch release combined with NGC~25.03, in which the underlying \texttt{libnvrtc} defect had been fixed.

\subsubsection{Power management}
Power management emerged as a further non-obvious but impactful constraint. For LLM workloads, which are predominantly compute-bound rather than memory-bandwidth–bound, the GH200 modules are typically power-limited rather than thermally limited. By default, a substantial fraction of the power budget is allocated to GPU memory. Enabling NVIDIA’s vBoost feature rebalanced this budget by reducing memory frequency and increasing GPU core frequencies while remaining within the same power envelope. Since LLM training benefits more from higher compute throughput than from peak memory bandwidth, this shift resulted in measurable gains in tokens per second without compromising stability. To operationalise this capability, vBoost was exposed via a custom Slurm option, allowing it to be selectively enabled for suitable workloads.

\subsection{The data bottleneck}
\label{subsec:journey__data_bottleneck}

Training Apertus at scale revealed significant sensitivity of throughput to storage behaviour.

\subsubsection{Storage and Filesystem I/O Challenges}
Early production runs showed substantial run-to-run variation in tokens per second despite identical model and parallelisation settings. Analysis of application and system metrics identified I/O interference on the shared global filesystem as the dominant residual source of variability, with unrelated workloads causing transient bandwidth and metadata slowdowns.

Different phases of the training workflow impose distinct storage access patterns: dataloaders, tokenizer outputs, and runtime caches rely on many small, latency- and IOPS-sensitive reads, whereas checkpointing is largely bandwidth-dominated. Treating these heterogeneous patterns uniformly on a shared filesystem amplified contention and increased susceptibility to external noise.

To mitigate this, data placement into storage was specialised by access pattern. Datasets, dataloader state, and runtime caches were migrated to SSD-backed, high-IOPS storage (\autoref{sec:infra_and_platforms}), while large sequential workloads were redirected to capacity-oriented tiers. This separation matched workloads to the storage system’s strengths and greatly reduced I/O-induced variability. After migration, tokens-per-second per GPU became markedly more stable.

At scale, contention extended beyond raw bandwidth. Concurrent access from thousands of ranks stressed Lustre and created hotspots when many processes hit the same files. File striping distributed I/O across multiple OSTs, improving load balance and throughput under high concurrency. It was especially important for large (approx. 20 GB) container images, where poor striping caused slow job start-up.

Additional filesystem pressure arose from software-generated artifacts. Triton JIT kernel caches were initially placed on the distributed filesystem; under concurrent access from thousands of ranks, this led to cache contention and race conditions, occasionally resulting in crashes. Relocating these caches to node-local memory eliminated contention entirely and improved overall job stability. Collectively, these measures transformed the storage subsystem from a dominant source of variability into a largely predictable component.

\subsubsection{Checkpointing, Restart Strategies, and Fault Tolerance}

Checkpointing was co-optimised with storage behaviour and scheduling semantics. Because checkpoint files consist of large, sequential writes with sustained bandwidth demand, they were directed to high-capacity HDD tiers rather than latency-optimised SSD storage, avoiding interference with IOPS-sensitive data paths while providing sufficient throughput and capacity. Checkpoints were written asynchronously \cite{pt_async_checkpointing_2025} so that training could continue during the long write operation; nevertheless, a small but measurable throughput dip was still observed while background writes were in progress.

Checkpoints were emitted every 250 iterations, a cadence derived using the Young–Daly formula \cite{apertus_report_2025}, which balances checkpointing overhead with the expected mean time between failures minimising re-computation in the event of a fault.

Fault tolerance was further strengthened through scheduler-aware orchestration. Training runs were chained using Slurm’s \texttt{--dependency=singleton} mechanism, ensuring that only one instance of a given training job could execute at a time and preventing accidental overlap following failures or requeues. Slurm’s \texttt{--signal} option notified jobs shortly before wall-time expiration, allowing a final checkpoint and clean termination.

Together, access-pattern-aware storage placement, asynchronous execution, and scheduler-integrated restart logic formed the fault-tolerance strategy. This combination limited wasted computation under realistic failure rates and avoided instabilities from excessive or poorly timed checkpoint I/O.

\subsection{The Communication Wall}
\label{subsec:journey__comm_wall}

\begin{table}[t]
\centering
\caption{Layered Apertus software and components stack}
\label{tab:apertus_layered_stack}
\begin{tabular}{p{1.78cm} p{1.95cm} p{3.85cm}}
\toprule
\textbf{Layer} & \textbf{Component} & \textbf{Role} \\
\midrule
Application   & Apertus            & The Swiss LLM application \\
LM Framework  & Megatron           & LLM training framework \\
DL Framework  & PyTorch            & Deep learning framework \\
GPU Runtime   & CUDA               & GPU execution and memory ops \\
Collectives   & NCCL               & DL collective communication \\
Network Bind  & AWS OFI Plugin     & NCCL--libfabric integration \\
Network API   & libfabric          & High-performance network API \\
Interconnect  & Slingshot-11       & HPE high-speed network \\
Storage       & Lustre             & Distributed file systems \\
\bottomrule
\end{tabular}
\end{table}

At large scale, LLM training on modern HPC systems is limited less by peak floating-point throughput than by the efficiency of the communication and integration layers that unify thousands of accelerators into a single job. In the Apertus campaign, this \emph{communication wall} became the main bottleneck once correctness and stability were established. \autoref{tab:apertus_layered_stack} illustrates the complexity of the components stack.

Application and system metrics analysis revealed several interacting bottlenecks spanning model parallelism, collective communication, and network integration. Profiling showed many small NCCL collectives during communication-heavy training phases. Although NCCL’s all-reduce, all-gather, and reduce-scatter are optimized, each call has fixed latency; frequent fine-grained collectives cause these latencies to accumulate and hurt scaling at large GPU counts, even when bandwidth is sufficient.

Increasing the Distributed Data Parallel (DDP) bucket size \cite{megatron_parallelism} in Megatron-LM mitigated this by fusing many small gradient exchanges into fewer, larger collectives, amortizing per-call latency and improving network utilization. Traces showed consolidated NCCL kernels, reduced synchronization overhead, and better compute–communication overlap. 

In parallel, two model-level modifications were required to scale reliably to 4096 GPUs. First, a Megatron-LM issue causing delayed computation of embedding gradients was removed, eliminating a late-stage increase in gradient norms that destabilised large-scale pre-training. Second, virtual pipeline parallelism \cite{megatron_parallelism} was increased from two to five layers per virtual stage within model-parallel groups. This increased pipeline concurrency and throughput, at the cost of higher communication volume, making efficient collective handling even more critical. 

\begin{figure*}[ht]
  \centering
  \includegraphics[width=0.7\textwidth]{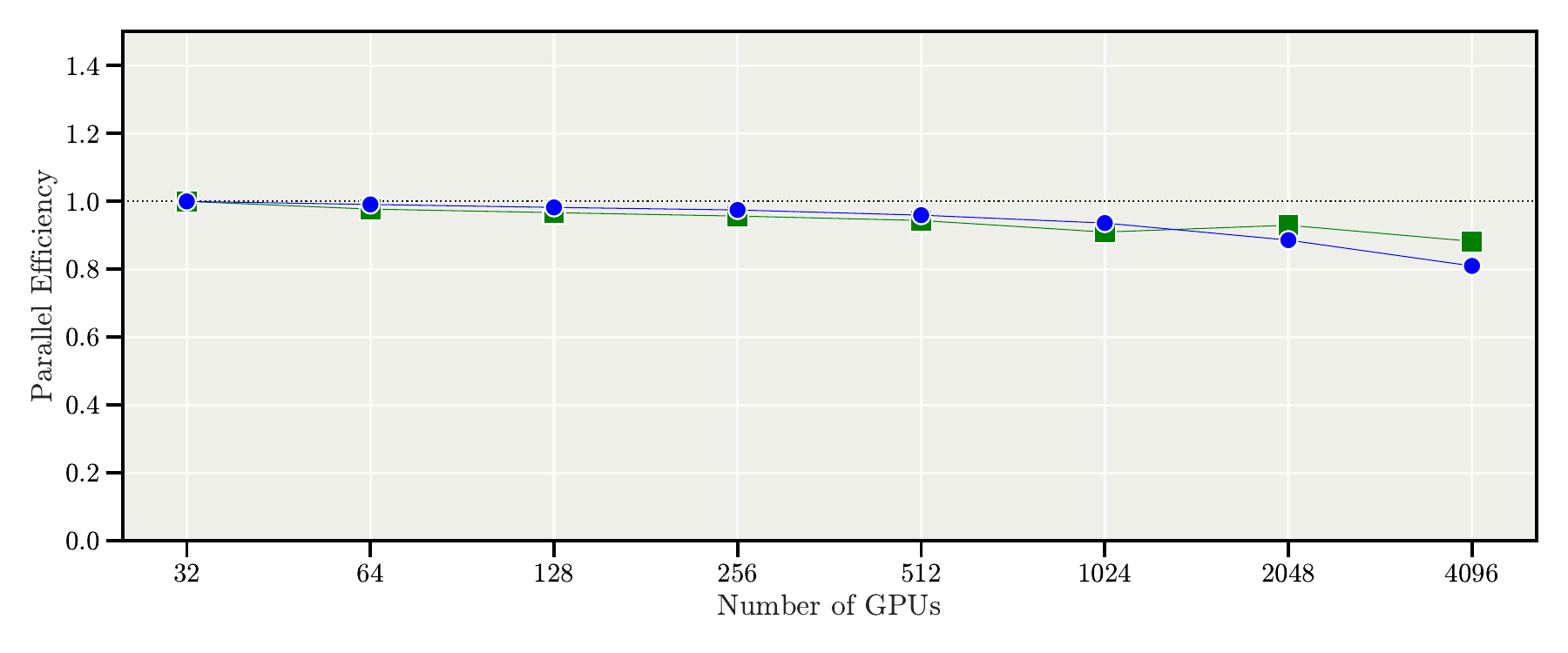}
  \caption{\textbf{Scaling of the Apertus 70B model}. Strong scaling parallel efficiency is shown with blue circles. Weak scaling parallel efficiency is shown with green squares. The global batch size was held constant at 16.8 M tokens for the strong scaling while the global batch size varies from 0.13 M to 16.8 M tokens with increasing GPU count for weak scaling. The figure is reproduced from the Apertus Technical Report for clarity\cite{apertus_report_2025}.}
  \label{fig:apertus_scaling}
\end{figure*}

After applying these model, communication, and networking optimisations, strong and weak scaling experiments for the 70B model across 32 to 4096 GPUs demonstrated sustained performance of approximately 723 Tokens/second/GPU, with around 80\% strong-scaling efficiency at the largest scale, as shown in \autoref{fig:apertus_scaling}. Achieving this efficiency required coordinated tuning across software layers and careful exploitation of low-level system features, underscoring how sensitive large-scale LLM training remains to communication patterns, network integration, and power-allocation choices.

\subsection{Stability and the long run}
\label{subsec:journey__stability}

Once baseline correctness, scaling behaviour, and peak throughput had been established, the primary challenge for Apertus' operational support shifted toward sustaining stable execution over weeks to months at scales of up to 4096 GPUs. At this point, stability became less a question of isolated fixes and more a system-level quality.

Many of the issues described above, runtime compilation failures, communication stack sensitivity, unified-memory side effects, and storage interference, were recurring risks whose impact grew with scale and runtime. Operational focus shifted to continuous awareness: detecting early degradation, distinguishing normal variability from emerging failures, and intervening before major compute time was lost.

To support this, scheduler features such as extended wall times, pre-termination signals, and checkpoint-aware orchestration were aligned with long-running training job configuration, enabling controlled shutdowns and predictable recovery. In parallel, continuous monitoring pipelines combined progress indicators from application logs with selected system telemetry, helping engineers interpret throughput trends and correlate anomalies with underlying infrastructure effects.

These practices were initially pragmatic responses to the demands of sustained production runs, but they exposed common patterns that could not be efficiently addressed through ad-hoc, campaign-specific solutions. The experience of operating Apertus over the long run thus provided the motivation and empirical grounding for the platform-level capabilities described in the following section, where operational insight is increasingly encoded into reusable services rather than relying on manual interpretation and intervention.

At the organisational level, a set of tools was introduced to help engineering teams prioritise work. Rapid debugging scripts, along with Key Performance Indicators (KPIs) displayed on kiosk dashboards around offices, all tailored around the Apertus campaign, provided a common view over system health and training progress relative to project deadlines, enabling faster decisions at engineering team level.

\subsection{Insights for Evolving the HPC Service Ecosystem}
\label{subsec:journey__evolving_HPC_services}

The Apertus pre-training campaign on Alps provided a real, large-scale and continuous, system-level test of the architectural choices underpinning the machine. Its duration, intensity, and workflow diversity surfaced limitations across networking, storage, node health, and observability that are difficult to uncover through synthetic benchmarks alone. As a result, several design directions for the ML Platform \cite{CUG_2025_evolving_hpc_for_ml} were developed or profoundly reshaped by the experience. The sections below summarise these directions and outline how they aim to improve and complete services on Alps for future large-scale LLM campaigns.

\subsubsection{Saturation Scorers}
These tools are being developed to organise the expert knowledge needed to analyse large-scale ML training workloads and make it more accessible to users. Diagnostics are essential for evaluating whether hardware is being used effectively, but their complexity and the time required make it difficult to run them regularly. These saturation scorers condense diverse hardware metrics into compact, digestible signals to support initial assessments. Current prototypes target GPU saturation, run as lightweight wrappers around the user application.

The term \emph{scorer} reflects their purpose: for less experienced users, they provide approximate but interpretable indicators of hardware saturation; for experts, they offer a simpler, \mbox{first-pass} alternative to in-depth analysis. Unlike application-level surrogates, such as tokens per second, these scores incorporate hardware-specific upper bounds.

Targeted at development and early scaling, saturation scorers provide fast, low-overhead feedback during iterative model and configuration changes. They help identify underutilization or misconfiguration before expensive large-scale runs and complement, rather than replace, detailed profiling and expert performance studies required for production tuning.

\subsubsection{Catalogues of Data Products}

Although sporadic, sudden throughput degradations during the Apertus training campaign exposed a structural vulnerability: effective triage depended on rapidly narrowing the space of plausible causes, yet this often requires collecting new data and assembling disparate datasets under significant time pressure. As publication deadlines approached, such incidents became acute organisational stressors, with engineering teams needing to re-establish situational awareness quickly. Accelerating this initial phase, where hypotheses are generated and quickly ruled out or refined, became a key operational need.

This motivated the development of structured \emph{catalogues of data products}: curated, ready-to-use collections of system telemetry, application metrics, ranks and nodes topology information, and other relevant data points. Their effectiveness depends not only on the breadth, completeness, and low-latency availability of these datasets, but equally on the quality of the interactive user interfaces that expose them.

To support refined hypothesis analysis, whether correlating temperature outliers with throughput drops or identifying network or file-system anomalies, these catalogues prioritise data \emph{usability}. Interactive views tailored to the underlying components provide filtering, colouring, cross-linking, and visualisation modes that reflect the structure and semantics of the data at hand. In doing so, they enable engineers to rapidly test root-cause hypotheses, discard implausible ones, and converge on actionable diagnoses.

In contrast to the saturation scorers used in development, these catalogues target the operational phase, where rapid situational awareness and diagnosis speed are critical.

\subsubsection{Node Vetting Tools for Early Aborts}

At large node counts, subtle node-state heterogeneity, such as thermal outliers or overlooked driver misalignment across node partitions, becomes increasingly likely to affect an allocation and can prevent jobs from starting cleanly, reaching stable performance, or cause frequent restarts in-between checkpoints.

To mitigate this, CSCS is developing centralized node-vetting and early-abort tools for end users based on workload-informed thresholds. Users can select lightweight tests relevant to their workload, which are then executed within the same allocation immediately before the application. Allocations are terminated early if inconsistent or suspicious node behaviour is detected, avoiding the waste of large GPU-hour budgets on runs that would not achieve stable efficiency or reach the next checkpoint iteration.

Providing this opt-in capability as a shared service removes the need for ad-hoc, project-local vetting solutions, which we observed to emerge independently across user projects, and enables systematic feedback to CSCS operation teams. While a temporary workaround rather than a substitute for inherent system reliability, it serves as an interim measure to improve users productivity as the platform matures.

\section{Fine tuning}

Fine-tuning uses the capabilities learned during pre-training to align model behavior, safety, and preferences with project goals. Of the several post-training stages involved, we discuss two here due to our more direct involvement.

\subsection{Supervised Fine-Tuning (SFT) Activities}

For the SFT stage, users prepared a large, high-quality instruction dataset together with iterative validation pipelines for safety-related filtering, encompassing licensing checks, decontamination, and ideological-sensitivity filtering aligned with user-defined safety and quality goals \cite[\S4.1.3--4.2]{apertus_report_2025}.

A key requirement was a reliable automated classifier to label the full dataset. Users first collected human ground-truth annotations on a representative subset, not to label the corpus exhaustively, but to evaluate and calibrate candidate models. After testing several classifiers on these volunteer-labeled samples, a Qwen3--32B-based classifier was selected \cite[\S4.1.4]{apertus_report_2025}. Once validated, this classifier was used to label the complete dataset.

CSCS enabled these workflows by providing the computational and data infrastructure required for high-throughput filtering, repeated classifier evaluation, and iterative SFT runs. The same infrastructure hosted a web-based application for the volunteer labelling task directly on the hosted datasets, allowing human annotation, classifier benchmarking, and model training to proceed in parallel without copying datasets elsewhere or workflow fragmentation. This integration was important for supporting rapid iteration across data curation and post-training activities.

\subsection{Reinforcement Learning}
Reinforcement Learning differs substantially from the earlier training stages due to its \emph{online} nature. Unlike pre-training or supervised fine-tuning, RL requires serving and evaluating multiple model instances in parallel, which greatly increases system complexity, especially in memory use, inter-process communication, and runtime stability.

At CSCS, we enabled large-scale execution of user-defined RL workflows on Alps, helping users address challenges in software maturity, memory management, and filesystem usage.

\subsubsection{Immature Software Stack}

The adopted software stack uses rapidly evolving frameworks such as VeRL, SGLang, Transformers, Triton, and PyTorch. These work well for research-scale experiments, but have seen limited use on large HPC systems. When scaling beyond thousands of GPUs, we faced issues stemming from cross-framework interoperability, implicit assumptions about cluster size, and the lack of production-hardened robustness.

Producing a stable, reproducible container image emerged as a significant major bottleneck. Ensuring stack-wide compatibility demanded many iterations to find consistent framework and patch combinations. Often, fixing runtime failures required unreleased patches or active pull requests from upstream repositories, further complicating reproducibility.

Iteration and validation were also limited by long turnaround times: building a single multi-stage container image could take 8–10 hours, hindering debugging and experimentation. This iteration cost led to a dedicated CI pipeline to automate image construction, testing, and validation, improving development speed and reducing configuration drift.

In several extreme cases, source-level intervention was necessary because some components were not designed for large-scale HPC. Notably, we identified and fixed a race condition in VeRL where random port assignments collided across serving engine instances on the same node. Although the conflict probability was low at small scale (approximately 0.6\%), it rose to roughly 78\% on 256 nodes, making long-running RL jobs impractical without a fix.

In other cases, patches were needed to expose previously hard-coded configuration options, such as timeouts. Defaults suitable for small-scale deployments failed at large scale, where higher latency and system noise are unavoidable.

\subsubsection{Memory Management}
Careful tuning of RL-specific hyperparameters was essential to scale training to thousands of GPUs. The complex software stack caused high baseline memory usage, leaving little headroom for inference and training. To manage this, the model was tensor-parallelized across GPUs, and batch and micro-batch sizes were chosen to balance memory pressure against throughput.

This tuning enabled long, uninterrupted 18–24 hour training runs; yet poor memory configuration could still lead to out-of-memory errors.

\subsubsection{Filesystem Management}
Another key challenge was efficiently loading large-scale models. Apertus 70B is about 150~GB, and VeRL’s default behavior was to load the model separately on each GPU. At scale, this triggered thousands of concurrent reads of the same data, generating multiple terabytes of simultaneous I/O, far beyond the possibilities of our Lustre-based file systems.

We addressed this by loading the model once on rank~0, then redistributing it to all GPUs over the high-speed network. This sharply reduced filesystem load, decreased job startup time, and improved system stability during large-scale RL runs.

\section{Lessons Learnt}
\label{sec:lessons_learnt}

The core of Apertus campaign spanned roughly eight months and provides a concrete reference for planning comparable large-scale efforts. Preparatory activities began in 2024 and intensified in early 2025 with data preparation, environment bring-up, development, and initial validation. Large-scale pre-training was conducted from March to early August 2025, with the 70B model finalized first, followed by the 8B variant. Post-training, evaluation, and the public release of models and artifacts were completed in early September 2025.

Because the ML field is advancing so rapidly, research and design activities could continue until the start of each phase, often concurrently with the technical difficulties intrinsic to bringing new methods into reliable operational use, making accurate timeline estimates difficult to establish or maintain.

Beyond these nominal phases, the schedule was shaped by hard-to-predict external and organisational factors, including extraordinary environmental events affecting infrastructure (e.g., a storm-induced interruption of our lake-water cooling system) and competing institutional priorities such as Gordon Bell prize–related activities. The key lesson is that large-scale project plans must explicitly budget for such contingencies and treat timelines as adaptive.

The Apertus campaign showed that training a state-of-the-art foundation model on a general-purpose academic supercomputer is feasible but demands substantial, sustained engineering effort and expertise across the full system stack (\autoref{sec:challenges} to \autoref{sec:journey}).

Large-scale LLM training is fundamentally a systems problem. Stability and performance depend more on interactions between operating systems, GPU drivers, memory management, interconnect software, storage, and container runtimes than on model architecture. In particular, defects and subtle behaviours in heterogeneous and unified memory management (\autoref{subsec:journey__sandbox_to_prod}) led to unpredictable GPU memory availability and kernel instability, with impact comparable to traditional I/O and communication bottlenecks.

At scale, data movement remains a primary limiting factor (\autoref{subsec:journey__data_bottleneck}). After core functionality was in place, residual throughput variability was largely driven by storage interference and collective communication inefficiencies (\autoref{subsec:journey__data_bottleneck} to \autoref{subsec:journey__comm_wall}). Stable performance required deliberate alignment of access patterns with storage tiers and reduction of fine-grained collectives.

Sustained, long-running training exposed the need for operational co-design. Multi-month executions amplified rare failure patterns and shifted critical effort toward evolving traditional HPC services (\autoref{subsec:journey__stability}, \autoref{subsec:journey__evolving_HPC_services}, \autoref{sec:future_works})

Platform elasticity proved to be a decisive enabler. The vCluster technology \cite{Alps_aVersatileRI} allowed CSCS to dynamically rebalance resources, temporarily increasing GPU capacity to meet release deadlines or guaranteeing minimum forward progress when other high-priority workloads occupied the majority of the system. This flexibility greatly supported operating a deadline-driven AI workload on shared national infrastructure.

Overall, the Apertus training strengthened CSCS in-house expertise and the robustness of the ML Platform. It also demonstrated that large-scale ML on emerging architectures requires sustained investment and close collaboration between CSCS engineers, hardware and software vendors (NVIDIA, HPE), and the SwissAI research teams.

\section{Future Works}
\label{sec:future_works}

The lessons captured in this report mark a starting point rather than a conclusion. The Apertus campaign yielded concrete operational insights that now guide the evolution of the ML Platform \cite{CUG_2025_evolving_hpc_for_ml} toward improved usability, robustness, and productivity. Near-term priorities include further maturation of the container engine with the transition from \texttt{enroot} to \texttt{Podman}; the development of tools for simplified performance assessment during both development and operational phases; and the provision of shared, project-centric mechanisms to assess node readiness and allocation quality ahead of large-scale executions. 

In parallel, the platform is expanding to support heterogeneous execution models, including inference workloads and ancillary services that demand proximity to the compute fabric, driven by emerging use-cases such as RAG and agentic applications. Ongoing work also targets tighter integration of inference and fine-tuning workflows, alongside improvements in data and storage management to address the growing complexity of datasets, preprocessed artifacts, and large numbers of checkpoints produced by concurrent post-training activities. These efforts are intended to deliver a better experience compared to using ad-hoc directory structures. Performance evaluations are underway to anticipate future non-text workloads and emerging architectures such as Mixture-of-Experts, enabling a shift from ad hoc, campaign-based efforts to a durable ML infrastructure where improvements from a campaign are consolidated into shared platform capabilities for the broader user community.

\section{Acknowledgements}
The core scientific activities described here were conducted by colleagues within the Swiss AI Initiative, and we refer to the \textit{Apertus} scientific report \cite{apertus_report_2025} for an in-depth discussion and a complete list of contributors to the project. The training and fine-tuning recipes developed within the Swiss AI Initiative, including the Megatron-LM fork on which Apertus is based are publicly available at \url{https://github.com/swiss-ai}.

We gratefully acknowledge the expertise, collaboration, and sustained engagement of all Swiss AI Initiative colleagues, and the essential contributions of our collaborators at HPE and NVIDIA, whose technical support and joint problem-solving were instrumental to the success of this effort.

The Apertus model itself has been used to improve the clarity and readability of this document.

\bibliographystyle{IEEEtran}
\bibliography{resources}

\end{document}